\documentstyle[epsf]{mn}

\input epsf

\title
[Large primordial GW background in a class of BSI $\Lambda$CDM models]
{Large primordial gravitational wave background
in a class of BSI $\Lambda$CDM models}

\author
[J.~Lesgourgues, D.~Polarski and A.~A.~Starobinsky]
{J. Lesgourgues$^{1}$, D. Polarski$^{1,2}$
and A. A. Starobinsky$^a$\\
$^1$~{\it Laboratoire de Math\'ematiques et Physique Th\'eorique,} \\
{\it UPRES-A 6083 CNRS, Universit\'e de Tours,}\\
{\it Parc de Grandmont, F-37200 Tours (France)}\\
$^2$~{\it D\'epartement d'Astrophysique Relativiste et de Cosmologie},\\
{\it Observatoire de Paris-Meudon, 92195 Meudon cedex (France)}\\
$^a$~{\it Landau Institute for Theoretical Physics}, \\
{\it Kosygina St. 2, Moscow 117334 (Russia)}}

\date{20 January 1999}

\begin{document}

\maketitle

\begin{abstract}
We consider the primordial gravitational wave (GW) background in a class of
spatially-flat inflationary cosmological models with cold dark matter (CDM),
a cosmological constant, and a broken-scale-invariant (BSI) steplike
primordial (initial) spectrum of adiabatic perturbations produced in an
exactly solvable inflationary model where the inflaton potential has a rapid
change of its first derivative at some point.
In contrast to inflationary models with a scale-free initial spectrum, these
models may have a GW power spectrum whose amplitude (though not its shape)
is arbitrary for fixed amplitude and shape of the adiabatic perturbations 
power spectrum. In the presence of a positive cosmological constant,
the models investigated here possess the striking property that a significant 
part of the
large-angle CMB temperature anisotropy observed in the COBE experiment
is due to primordial GW. Confronting them with existing
observational data on CMB angular temperature fluctuations, galaxy
clustering and peculiar velocities of galaxies, we find that for the best
parameter values $\Omega_{\Lambda}\approx 0.7$ and $h\approx 0.7$, the GW
contribution to the CMB anisotropy can be as large as that of the
scalar fluctuations.
\end{abstract}


\section{INTRODUCTION}

There seems to be increasing observational evidence in support of the
inflationary paradigm (see the review in Linde 1990; Kolb \& Turner 1990;
Lyth \& Riotto 1998), which offers an elegant solution to some
of the outstanding problems of standard Big Bang cosmology. During
inflation, primordial quantum fluctuations
(Hawking 1982, Starobinsky 1982, Guth \& Pi 1982) of some scalar field(s)
(inflaton(s)) are produced, which eventually form galaxies,
clusters of galaxies and the large-scale structure
of the Universe through gravitational instability.

However, the simplest inflationary CDM model in flat space 
($\Omega = 1$) and with an approximately flat
($n_S\approx 1$) initial spectrum of adiabatic perturbations does not
agree with observational
data because it has too much power on small scales when normalized
to the COBE data at large scales. 
One of the simplest and most elegant ways to reach agreement with
observations is to add a positive cosmological
constant to dustlike CDM. On one hand, this improves pure CDM models a great
deal; on the other hand, the cosmological constant itself is viable
in the presence of cold dark matter only (as was emphasized e.g. in
Kofman \& Starobinsky 1985). Now, the $\Lambda$CDM model is perhaps the
most promising CDM variant (Bagla, Padmanabhan \& Narlikar 1996;
Ostriker \& Steinhardt 1995). 
A recent argument in favour of $\Omega < 1,~(\Omega\equiv 1-\Omega_{\Lambda})$
(and $\Omega \simeq (0.2-0.4))$\ follows from the evolution
of rich galaxy clusters (Bahcall, Fan \& Cen 1997;
Fan, Bahcall \& Cen 1997; Eke et al., 1998).
The most recent data on SNe Ia events at large redshifts also strongly
favour $\Omega < 1$ (Perlmutter et al. 1998; Garnavich et al.
1998; Riess et al. 1998), while the latest CMB constraints
seem inconsistent with these low $\Omega$ values in case $\Omega_\Lambda=0$
(Lineweaver 1998). Finally, galaxy abundance at large redshifts also
disfavours open models (Peacock 1998),
leaving the $\Lambda$CDM model with $n_S \approx 1$ as the most
successful model for structure formation.

On the other hand, the observational fact that the smoothed slope
of the initial spectrum $n_S$ cannot be significantly
different from 1 does not exclude
the possibility of local strong deviations from scale invariance, i.e.,
steps and/or spikes in the initial spectrum. Of course, such a behaviour
is not a typical one, and may be expected to occur only in exceptional points 
in Fourier space (if at all). At present, one scale 
$k \approx 0.05 h$Mpc$^{-1}$ is a candidate for this role, since there
exists an evidence for a peculiar behaviour (in the form of a peak or a sharp
change in $n_S$) in the Fourier power spectrum of rich Abell-ACO clusters
of galaxies around this point
(Einasto et al. 1997a, 1997b, 1997c; Retzlaff et al. 1998). 
This evidence is, certainly, inconclusive, especially since there is no
peak at $k \approx 0.05 h$Mpc$^{-1}$ in the power spectra of both
APM clusters (which are generally less rich than Abell-ACO clusters)
and APM galaxies (Tadros, Efstathiou \& Dalton 1998), 
though a break in $n_S$ may exist in the latter spectrum too
(Gazta\~naga \& Baugh 1998). Note that
the attractive possibility to explain this feature by Sakharov (acoustic)
oscillations fails (Atrio-Barandela et al. 1997, Eisenstein et al. 1997).
Hence, if confirmed in the future, this feature should originate in the
initial perturbation spectrum itself.

Therefore, there exist serious reasons to continue the investigation of BSI 
CDM models. Then, it is natural to fit the
arbitrary wavenumber of the singularity appearing in these models to
$k =0.05 h$Mpc$^{-1}$. From the theoretical point of view, inflationary
models with BSI spectra having such a characteristic scale naturally
appear as a result of a phase transition approximately 60 e-folds before
the end of inflation (Kofman \& Linde 1986; Adams, Ross \& Sarkar 1997;
see also Starobinsky 1998). However, it is important that inflation cannot 
produce very sharp features in the primordial spectrum. In particular,
an exact $\theta$-function-like step (which would result from a naive
application of the slow-roll expressions to the case in which the first 
derivative of an inflaton potential $V(\varphi)$ sharply changes near some
point $\varphi=\varphi_0$) is not permitted; what appears instead follows
from the exact solution found in Starobinsky (1992).

Summarizing, if we believe {\em both} 
in the present cosmological constant and in the
cluster data exhibiting a preferred scale in the perturbation
spectrum, we have to consider the $\Lambda$CDM model with a BSI initial
spectrum. Remarkably, a detailed comparison with the bulk of all existing 
observational data shows that the $\Lambda$CDM model plus the above
mentioned BSI spectrum gains much from the cosmological constant too
(Lesgourgues, Polarski \& Starobinsky 1998, hereafter Paper I). 
Not only does the
inclusion of $\Lambda > 0$ enlarges the allowed region of cosmological
parameters $\Omega$ and $H_0$, but it also permits the new possibility of an 
{\it inverted} step (i.e. more power on small scales) in the initial spectrum.
Also, this model avoids the main problem found in a model of double inflation
(Lesgourgues \& Polarski 1997) for which the Doppler
peak turns out to be
low even for those values of the parameters with an acceptable matter power
spectrum $P(k)$.
However, in Paper I, the free parameters of the model were chosen in such
a way that the primordial GW background generated during inflation in
addition to adiabatic (scalar) perturbations was too small to
contribute significantly to the observed large-angle CMB temperature
anisotropy. Now we want to investigate if it is possible to obtain
a large GW contribution to the temperature anisotropy $\Delta T/T$ in this
model.

Historically, the primordial GW background generated from quantum
vacuum metric fluctuations during inflation was the first
observational prediction of inflation: its spectrum was first calculated in
Starobinsky (1979) even before the first viable models of inflation
were constructed, and the CMB temperature anisotropy for the multipoles
$l=2,3$ produced by this background was found by Rubakov, Sazhin \&
Veryaskin (1982); see Lyth \& Riotto (1998) for references on numerous
subsequent papers. 
The GW and adiabatic contributions to $\Delta T/T$ cannot be separated if 
only the latter quantity is measured, so the quantity
of practical interest is actually $\sqrt{1+{C^T\over C^S}} - 1$, i.e.,
the relative excess of the observed rms value of $\Delta T/T$ over the
rms value calculated under the assumption that there is no GW contribution
at all. However, measurement of the CMB polarization provides a unique
opportunity to distinguish both these contributions and to prove the
existence of a primordial GW background directly.
In the case of the simplest inflationary models with scale-free
spectra, $C^T_{10}/C^S_{10}$ can be very small, 
e.g., in case of the ``new'' inflation
or the $R+R^2$ inflation (where $R$ is the Ricci scalar). On the other
hand, for chaotic inflationary
models with a power-law inflaton potential $V(\varphi)$, $C^T_{10}/C^S_{10}$ is
comparable with unity, though still rather small, e.g., $C^T_{10}/C^S_{10}
\approx 0.2$ for the quartic potential.
Another example is provided by power-law inflation with an exponential 
potential $V(\varphi)$ leading to a tilted CDM model. 
However, such models did not prove to be successful in explaining the height
of the first accoustic (Doppler) peak for $n_S< 0.9$, while $n_S > 0.9$
results in $(C^T_{10}/C^S_{10}) < 0.5$.
Note that inflationary models having non-negligible $C^T_{10}/C^S_{10}$ 
may be also
characterized by the condition that the total variation of the inflaton
field during the period corresponding to present scales in the range
$(10-10^4)h^{-1}$Mpc is not negligible compared to the Planck mass (Lyth 1997).

The obstacle for having a sufficiently large $C^T_{10}/C^S_{10}$ in the 
case of a
scale-free initial perturbation spectrum lies in the observational fact
that the bandpower $l(l+1)C_l$ grows almost by an order of magnitude when
$l$ changes from 10 to 200 - 300 (see Table 1 below). This is even larger
than what is expected in the
standard $n_S=1$ CDM model without GW and a cosmological constant, while
the presence of a significant GW contribution to $\Delta T/T$ on large angles 
should manifest itself in a decrease of the height of the first Doppler
peak relative to the normalization at $l=10$. In other words, a high
first Doppler peak is an argument against significant primordial
GW background in case of scale-free initial perturbation spectra. Note that
even using old CMB data from the Saskatoon 94 and South Pole 94 experiments 
referring to $l=(70\pm 20)$, it was already possible to reach the conclusion
that $(C^T_{10}/C^S_{10})<0.7$ with 
$97.5\%$ probability in the cases of chaotic
and power-law inflation with $h=0.5$ and $\Omega=1$ (Markevich \&
Starobinsky 1996). Now, with the whole set of data presented in Table 1,
this upper limit becomes much lower. The existence of a positive
cosmological constant only slightly relaxes this argument, so that
we still get $(C^T_{10}/C^S_{10}) \le 0.15$ in this case (see Sec. 3.2
below, the case $p=1$).

Thus, at present, the only possibility to have a significant primordial GW
background is to require some kind of significantly non-scale-invariant
initial density power spectrum. The first attempt in this direction was
performed by Lukash \& Mikheeva (1996, 1998) who considered the case of
the inflaton potential $V(\varphi)=V_0+{m^2\varphi^2\over 2}$ with
$m^2$ not small as compared to $H_0^2\equiv 8\pi GV_0/3$. However, this
model has a ``blue'' initial spectrum with $n_S>1$ at small scales, and
faces serious problems regarding its excess of power at scales
smaller than $8h^{-1}$Mpc if $n_S\ge 1.3$. On the other hand, if the parameters
of the model are taken in such a way that the asymptotic slope 
satisfies $n_S<1.3$,
then $(C^T_{10}/C^S_{10})<0.5$. So, it appears that in order to get 
$(C^T_{10}/C^S_{10})=1$
or more, one has to take a BSI spectrum with a significantly 
sharper feature in it. 
This gives one more reason to investigate the model
considered in Paper I with respect to the possible existence of a large 
primordial GW background.

Therefore, as in paper I, we suppose that the inflaton potential derivative,
$V'(\varphi)$, has a rapid change from $A_+$ to $A_-$, when $\varphi$ 
decreases, in a neighbourhood of
$\varphi_0$. This generates a steplike spectrum of
primordial adiabatic fluctuations, with a specific substructure.
In a vicinity of the step, the amplitude of large-scale and small-scale
plateaus is given by:
\begin{equation} \label{Phi}
k^3 \Phi^2(k)\equiv \frac{81}{50} \frac{H_0^6}{A_\pm^2}, \qquad
H_0^2=\frac{8\pi G}{3} V(\varphi_0),
\end{equation}
where $\Phi$ is the (peculiar) gravitational potential at the matter
dominated stage.
So, the amplitude of the step is given by $p\equiv A_-/A_+$, and we call $k_0$
the characteristic scale of the step (for details, see paper I).
In this model, it is still possible to fix freely the amount of
primordial GW for given $p$ and normalization,
since their initial spectrum near $k=k_0$ is given, for each
polarization state, by the standard expression:
\begin{equation}
k^3 (h_{mn}h^{mn})(k)=16\pi G H_0^2
\label{h}
\end{equation}
which does not depend
on $A_{\pm}$ in contrast to $k^3 \Phi^2(k)$ (here $m,n=1,2,3)$.
Without the inclusion of a cosmological constant, we would be forced to
consider the case $p>1$\ only, in order to increase power on large scales.
Since $\Lambda>0$\ already produces a desired excess of large-scale power,
we are now free to consider both cases $p>1$\ and $p<1$.

However, in order to confront the model with observational data, we need 
the expression for the primordial spectra far from the point $k=k_0$, up to
$|y|=|\ln (k/k_0)|\sim 5$. Since we adopt the natural assumption that
any deviation from the slow-roll regime requires some special origin (e.g.,
some kind of phase transition during inflation) and, thus, that it should be 
an exceptional phenomenon, we assume that both slow-roll conditions given
above are valid everywhere far from the point $k=k_0$ (before the end of
inflation, of course). Then standard expressions for the initial
spectra of adiabatic perturbations and GW are obtained in this region.
So, the full initial spectra for these perturbations follow from (\ref{Phi})
and (\ref{h}), respectively, by the substitution:
\begin{equation}
H_0^2\to H_k^2\equiv {8\pi GV_k\over 3}, \qquad
A_{\pm} \to {V'}_k
\label{spec}
\end{equation}
where the index $k$ means that the quantity is taken at the moment
of the first Hubble radius crossing ($k=aH$) during the inflationary stage.

Note that since the second derivatives of $V(\varphi)$ are not fixed by
Eq.(1), we may freely take $n_S(k)$ far from the break point
$k=k_0$. However, $|n_S - 1|$ should be small in this region due to the
two slow-roll conditions mentioned above. In paper I, we assumed
that the smooth part of $|n_S-1|$ is so small that it can be neglected
at all for $|\ln(k/k_0) |\le 5$, so the upper and lower plateaus of the scalar
spectrum may be considered as flat ones. This assumption is self-consistent
in the case of a negligible GW background. In our case, the situation is more
complicated. In the slow-roll regime we have the well-known
relation:
\begin{eqnarray}
{C^T_{10} \over C^S_{10} } \approx -5 n_T={5 \over 8\pi G}
\left({V'\over V}\right)_k^2, \nonumber \\
{d\varphi_k\over d\ln k}=-{1\over 8\pi G}
\left({V'\over V}\right)_k,
\label{nt}
\end{eqnarray}
where the latter expression implicitly defines $\varphi_k$ as a function
of $k$. The coefficient 5 appearing here is approximate, its exact value
depends on both $h$ and $\Omega$ (see Bunn, Liddle \& White 1996 for
a fitting expression at $l=14$, which can be easily transformed to $l=10$).
Further, small corrections proportional to $n_T$ and $(n_S-1)$ should be
also taken into account if one wants to make this coefficient even more 
accurate 
(see Lidsey et al. (1997) for a review). Note that if we neglect the existence 
of a radiation dominated stage in the past (which corresponds to the 
approximation $t_{eq} \ll t_{rec}$), it would be equal to 5.3 for $\Omega=1$,
$n_S=1$, $n_T=0$ (as was given in Polarski \& Starobinsky 1995).

Since we are interested in the case when $C^T_{10}/C^S_{10}$
is not small, we cannot assume that $n_T\approx 0$.
So, the amplitude of the GW background 
determines the slope of its power spectrum.
Moreover, the values of $n_T$ on the right and on the left
sides of the break point are different:
\begin{equation}
n_T(k>k_0)-n_T(k<k_0) = {A_+^2-A_-^2\over 8\pi GV_0^2}.
\label{break}
\end{equation}
Strictly speaking, this equation refers to the smooth part of
$n_T$ defined by differentiation of $H_k^2$ only; in addition, there is
some substructure near $k=k_0$. However, for our choice of $k_0$,
the only quantity of interest is $n_T(k)$ for $k=(0.01-0.5)~k_0$.

Therefore, we have to specify more accurately our initial spectra (or, 
equivalently, $V(\varphi)$). As a result, the BSI CDM model which we consider 
in this paper is actually different from the one with negligible GW studied
in Paper I (though they have the same structure at $k \approx k_0$).
We shall consider the two most representative cases.

1) $n_S\approx 1$ for both positive and negative large values of $\ln (k/k_0)$.
In the slow-roll regime, this corresponds to an inflaton potential
approximately proportional to $(\varphi - \varphi_{\pm})^{-2}$
for $\varphi > \varphi_0$ and $\varphi < \varphi_0$ respectively,
where $\varphi_{\pm}$ are two different constants. 
As discussed
above, the choice of the initial scalar spectrum with flat plateaus
at large and small $k$'s is very suitable for an explanation of both the 
small-scale observational data and the COBE results. Then, however,
we cannot assume $n_T(k)$ to be constant: it must satisfy
the equation (the ``consistency relation'' for single-field slow-roll
inflation) which takes the following form for $n_S=1$:
\begin{equation}
{dn_T\over d\ln k}=n_T^2
\end{equation}
(see, e.g., Eq.(14) in Polarski \& Starobinsky 1995 and the more general
discussion in the review by Lidsey et al. 1997). As expected,
this case leads to the largest possible values of $C^T_{10}/C^S_{10}$.

2) $n_S \approx 1$  for positive large values of  $\ln(k/k_0)$, but
$n_T=n_S-1= {\rm const} <0$ for negative large $\ln(k/k_0)$.
This corresponds to the exponential form of $V(\varphi)$ 
above $\varphi_0$. Here we may assume constant
slopes far from the break point $k=k_0$.

\section{CONFRONTATION WITH OBSERVATIONS}

We want now to compare our model with observations. We use experimental
data which constrain the matter power spectrum $P(k)$ on one hand and the
radiation power spectrum on the other hand. As we have already noted earlier
(Lesgourgues \& Polarski 1997) when dealing with a model of double inflation,
another model with broken
scale-invariant primordial spectrum, the constraints on both types of
fluctuations are essentially complementary, so that successfull confrontation
of one spectrum of fluctuations does not preclude bad results for the other
spectrum. One has to remember that the point of this analysis is to
investigate how large the amplitude of the produced GW can be while leaving
the model still in good agreement with observations. Hence, we must
adopt a strategy that enables us to answer this question with sufficient
accuracy while leaving the possibility to refine the analysis later on if
required by new observational evidence.

The primordial scalar spectra under investigation have four free parameters:
an overall normalization factor $Q_{10}$,
$p$, $k_0$, and $C^T_{10}/C^S_{10}$. Given these four parameters, and some
cosmological parameters, the scalar and (running) tensor tilts are completely
set, for each possibility $n_S=1$\ or $n_S=1+n_T$.
We choose the step location as well as the overall normalization fixed:
$k_0=0.016 \,h\,{\rm Mpc}^{-1}$, which corresponds to a bump in the present
matter power spectrum $P(k)$\ at $k=0.05 \,h\,{\rm Mpc}^{-1}$, while we take
$Q_{10}=18\mu K$ (Bennett et al. 1996).
The parameter $p$ is found by requiring that $\sigma_8=0.60\,\Omega^{-0.56}$
(White, Efstathiou \& Frenk 1993; see also Viana \& Liddle 1998), where
$\sigma_8$ is the variance of the total mass fluctuation in a sphere of
radius $8\,h^{-1}{\rm Mpc}$. Then, all models
are simultaneously normalized to COBE (large scales, scalar plus tensor
components) and to $\sigma_8$\ (small scales, scalar component only).
Note that $p$ is still a function of the remaining
three free parameters: the two cosmological parameters $h$ and
$\Omega_{\Lambda}$ on one hand, the inflationary parameter
$C_{10}^T/C_{10}^S$ on the other hand. In this way, the latter is singled
out as the only remaining inflationary free parameter.

We then compute numerically, using the fast Boltzmann code {\sc cmbfast} by
Seljak \& Zaldarriaga (1996), the matter power spectrum $P(k)$ and the CMB
power spectrum $C(\theta)$ for various values of the parameters 
$h,~\Omega_{\Lambda},~C_{10}^T/C_{10}^S$. 
We will find it interesting to state the results in
two-dimensional cuts of the parameter space for several given values of $h$.
This is particularly appropriate if we expect other observations to yield some
refined a priori knowledge of the value of $h$.

In order to constrain the matter power spectrum, we use:
\begin{itemize}
\item
peculiar velocities taken from the MARK III catalog (Willick et al. 1997) 
and the POTENT reconstruction (Bertschinger \& Dekel 1989; Kolatt \& Dekel 
1997) of the velocity field. Peculiar velocities have the advantage
that they probe all mass (and not just galaxies), but they have rather
large uncertainties. We use here the rms bulk velocity at
$R=50\,h^{-1}{\rm Mpc}$, with Gaussian smoothing radius
$R_s=12\,h^{-1}{\rm Mpc}$:
\begin{equation}
290 \leq V_{50} \leq  460~{\rm km~s^{-1}}~~~(1-\sigma~{\rm confidence ~level})
\end{equation}
in the absence of cosmic variance, and:
\begin{equation}
245 \leq V_{50} \leq 505~{\rm km~s^{-1}}
\end{equation}
when cosmic variance is taken into account.
As we will see, the lower bounds
implied by these results yield stringent constraints on our models.
\item
the STROMLO-APM redshift survey which gives a count-in-cells analysis of
large scale clustering. Like other redshift surveys, it probes the matter
perturbations well into the linear regime. Results are given in Loveday
et al. (1992) for cells of nine different sizes. We compare these
data points with the power spectrum (normalized to $\sigma_8=1.00$, and
convolved with the nine corresponding window functions) through a $\chi^2$\
analysis. Since we can vary 3 parameters, there are 6 degrees of freedom.
\item
the power spectrum of rich Abell galaxy clusters, consisting of 36
points taken from Einasto et al. (1997a). Since a few points on the
largest and smallest scales suffer from large uncertainties, we perform a
$\chi^2$\ analysis with only 30 points, corresponding to $0.023 \leq k \leq
0.232 \,h\,{\rm Mpc}^{-1}$.
As already said in the introduction, this spectrum exhibits a clear feature
with a bump at $k=0.05 \,h\,{\rm Mpc}^{-1}$ and $k_0$ is chosen so that the
matter power spectrum also exhibits a feature at the right scale.
Hence, we are testing here a theory with 4 free parameters,
and the $\chi^2$\ distribution has 26 degrees of freedom.
Strictly speaking, the Einasto et al. (1997a) data are not independent. 
Taking into account their correlation will decrease the effective number of 
degrees of freedom, however this does not noticeably change our conclusions. 
We do not assume any specific value for the biasing factor for these
data: for each set of parameters $(h,~\Omega_{\Lambda},~C_{10}^T/C_{10}^S)$, we
calculate and adopt the biasing factor yielding the smallest $\chi^2$.
In that sense, this test probes only the shape of the power spectrum.
\end{itemize}

In order to constrain the radiation power spectrum, we use:
\begin{itemize}
\item
the CMB anisotropy data on large and small angular scales found in
various experiments. We
use the bandpower estimates $\Delta T_l \pm \sigma$ given for an experiment
characterized by a window function $W_l$.
More precisely, we have:
\begin{equation}
\Delta T_l = \frac{\sigma_{obs}(T)}{\sqrt{I(W_l)}}~,
\end{equation}
where $\sigma_{obs}(T)$, the observed rms temperature fluctuation (which
differs from the theoretical value by the inclusion of the window function
$W_l$ characterizing the experiment), is given by:
\begin{equation}
\frac{\sigma^2_{obs}(T)}{T^2} =
\sum_{l=2}^{\infty}\frac{2l+1}{4\pi}C_l W_l~,
\end{equation}
and $I(W_l)$\ is defined as:
\begin{equation}
I(W_l)= \sum_{l=2}^{\infty}(l+\frac{1}{2})\frac{W_l}{l(l+1)}~.
\end{equation}
The bandpower estimate assumes a flat spectrum $l(l+1)C_l\propto C_2$ being
a good approximation over the width of the window function $W_l$ centered
around some ``effective'' multipole number $l_e$ defined through
$l_e \equiv I(lW_l)/I(W_l)$.
We compare the theoretical CMB power spectrum with 19 experimental points
given in Table 1, again through a $\chi^2$ analysis,
with 16 degrees of 
freedom\footnote{After this paper was submitted, the results of
two new important CMB experiments have been published. 
They confirm the data presented in Table 1, and, therefore, strenghten our
conclusions. The first experiment, QMAP (de Oliveira-Costa et al. 1998)
confirms the Saskatoon normalization at $\ell=90$, and the shape of the
Saskatoon data up to $l=150$. The second one, the NCP at OVRO
(Leitch et al. 1998) confirms the second CAT 1 point at $l\approx 600$.}.
The Saskatoon data provide an important constraint on any candidate model,
as they give a clue to the correct height of the first Doppler (or acoustic) 
peak. However these are given in Netterfield et al. (1997)
with a calibration error of $\pm~14\%$, which applies equally to all five
points. This remaining uncertainty
is treated as in Lineweaver \& Barbosa (1998a, 1998b): for each set of free
parameters $(h,~\Omega_{\Lambda},~C_{10}^T/C_{10}^S)$,
we make a preliminary $\chi^2$ analysis in order to find the best calibration.
Then, the Saskatoon data are treated on equal footing with other CMB data,
and a second $\chi^2$\ is computed with the 19 points.

\end{itemize}

\begin{table*}
\begin{minipage}{120mm}
\label{Table1}
\caption{Bandpower estimates used for the CMB $\chi^2$ test.}
\begin{tabular}{|l|l|r|l|} \hline
Experiment & $\Delta T_l \pm \sigma$ ($\mu$K) & $l_e$ & Reference \\
\hline
 Tenerife       & $32.5^{+10.1}_{-8.5}$ & 20 &Guti\'errez et al. (1997)\\
 South Pole 91  & $30.2^{+8.9}_{-5.5}$  & 60 & Gunderson et al. (1995)\\
 South Pole 94  & $36.3^{+13.6}_{-6.1}$ & 60 & Gunderson et al. (1995)\\
 Python         & $57.3^{+19.1}_{-13.6}$& 91 & Ruhl et al. (1995)\\
 ARGO 1         & $39.1^{+8.7}_{-8.7}$  & 95 & De Bernardis et al.
(1994)\\
 ARGO 2         & $46.8^{+9.5}_{-12.1}$ & 95 & Masi et al. (1996)\\
 MAX GUM        & $54.5^{+16.4}_{-10.9}$& 138 & Tanaka et al. (1996)\\
 MAX ID         & $46.3^{+21.8}_{-13.6}$& 138 & Tanaka et al. (1996)\\
 MAX SH         & $49.1^{+21.8}_{-16.4}$& 138 & Tanaka et al. (1996)\\
 MAX HR         & $32.7^{+10.9}_{-8.2}$ & 138 & Tanaka et al. (1996)\\
 MAX PH         & $51.8^{+19.1}_{-10.9}$& 138 & Tanaka et al. (1996)\\
 Saskatoon      & $49^{+8}_{-5}$  & 86 & Netterfield et al.
(1997)\\
 Saskatoon      & $69^{+7}_{-6}$  & 166 & Netterfield et al.
(1997)\\
 Saskatoon      & $85^{+10}_{-8}$ & 236  & Netterfield et al.
(1997)\\
 Saskatoon      & $86^{+12}_{-10}$ & 285 & Netterfield et al.
(1997)\\
 Saskatoon      & $69^{+19}_{-28}$ & 348 & Netterfield et al.
(1997)\\
 CAT 1          & $51.8^{+13.6}_{-13.6}$ & 396 & Scott et al. (1996)\\
 CAT 2          & $57.3^{+10.9}_{-13.6}$ & 396 & Baker et al. (1998)\\
 CAT 1          & $49.1^{+19.1}_{-13.7}$ & 607 & Scott et al. (1996)\\

\hline
\end{tabular}
\end{minipage}
\end{table*}

We chose to perform three separate $\chi^2$ analysis for the Stromlo-APM,
the CMB and the rich Abell cluster data.
In this way, we see clearly what is implied by each set of data separately.
We further avoid to put on the same footing observations of a very
different kind.

\section{RESULTS}

\begin{figure*}
\caption[]{Results of all the tests, in both case $n_S=1$\ ans $n_S=1+n_T$,
for a few two-dimensional cuts of the parameters space corresponding to
$h=0.5,0.6,0.7,0.8$. The preferred regions for each test are limited by
dotted curves in blue (bulk velocities), red (STROMLO-APM), black (cluster
distribution), and green (CMB anisotropy). Inside the former three regions,
we plot the contours associated with integer values of $\chi^2$.  }
\label{SIGW}
\epsfxsize=16cm
$$
\epsfbox{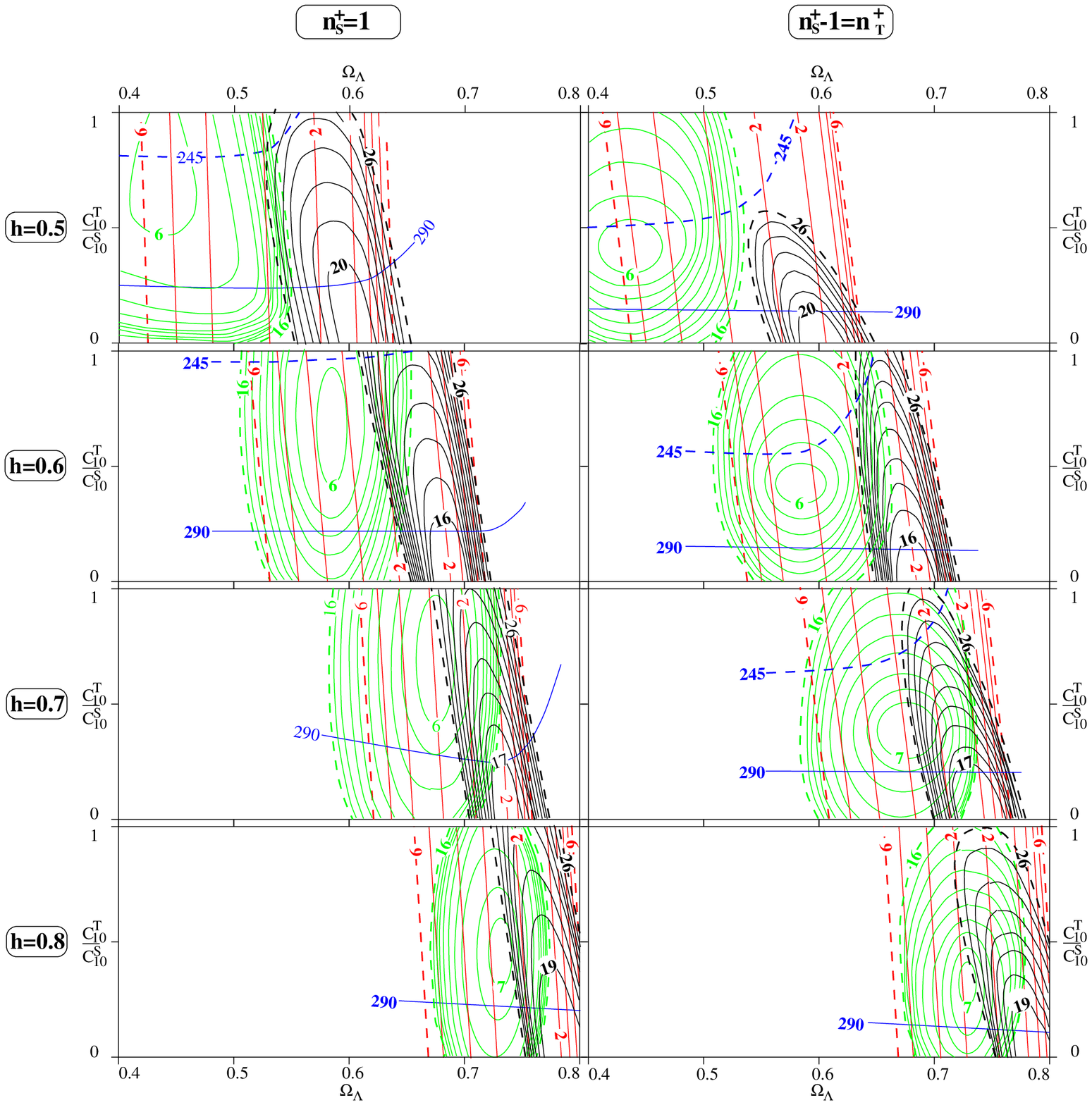}
$$
\end{figure*}

The result of all the tests are given in Figure \ref{SIGW},
in both case $n_S=1$\ ans $n_S=1+n_T$, for a few
two-dimensional cuts of the parameter
 space corresponding to
$h=0.5,0.6,0.7,0.8$. For peculiar velocities, we plot the curves
corresponding to the lower limit $V_{50}=245~{\rm km~s^{-1}}$
(the limit without cosmic variance, $V_{50}=290~{\rm km~s^{-1}}$, is also
given for completeness). For the three $\chi^2$\ tests, we plot contours
inside the preferred regions, in which $\chi^2$ is smaller or equal to the
number of degrees of freedom. Let us comment these results step by step.
Since the results for the $n_S=1$\ and $n_S=1+n_T$\ are quite similar,
we will discuss them simultaneously.

\subsection{Peculiar velocities and STROMLO-APM}

First, let us not consider the rich Abell galaxy clusters distribution.
In this case, the matter power
spectrum is constrained, on one hand, by bulk velocities, and on the other
hand, by the STROMLO-APM redshift survey. The bulk velocity test probes
the amplitude of $P(k)$\ at intermediate scales, corresponding to the step
in the primordial spectrum. This amplitude can vary significantly, even if
$\sigma_8$\ has been fixed on smaller scales, through the matter
transfer function {\it and} the primordial spectrum (indeed, a change in
$C_{10}^T/C_{10}^S$\ induce a change in $p$). Hence, this test depends on 
all three
parameters. One can see on Figure \ref{SIGW} that
for ($n_S=1$, $h\leq 0.6$), and for ($n_S=1+n_T$, $h\leq 0.8$),
this constraint cuts the preferred region in the
$C_{10}^T/C_{10}^S-\Omega_{\Lambda}$ parameter space
below the value $C_{10}^T/C_{10}^S=1$. 

On the other hand, the STROMLO-APM redshift survey
probes the shape of $P(k)$\ only at scales smaller than the step,
{\it i.e.} those scales given by the  high-$k$
plateau of the primordial spectrum.
Hence, it only depends on the matter transfer function,
not on $C_{10}^T/C_{10}^S$, and on Figure \ref{SIGW} it appears as a
constraint on $\Omega_\Lambda$\ only.

These two constraints define a wide preferred region in parameter space,
and it is crucial to include CMB anisotropy measurements in order to obtain
good predictions.

\subsection{CMB anisotropies}

The CMB $\chi^2$\ test selects preferred elliptic regions in the
$C_{10}^T/C_{10}^S-\Omega_{\Lambda}$\ planes,
which truly admit an extension up to $C_{10}^T/C_{10}^S=1$\ and even more.
This is due to the `inverted' step in the primordial spectrum,
which compensates for the loss of power in small-scale anisotropy usually
implied by a high tensor contribution.
The GW contribution even improves the model: in all cases plotted in
Figure \ref{SIGW},
the best values are obtained for $0.4<C_{10}^T/C_{10}^S<0.8$.

\begin{figure*}
\caption[]{On this plot, we coloured (yellow or gray)
the intersection of velocity, STROMLO-APM and CMB tests.
The intersection of all tests, including cluster ditribution, is the
bright yellow (or dark gray) region alone.
High values of $C_{10}^T/C_{10}^S$\ up to 1
are allowed, or even preferred.}
\label{SIGW2}
\epsfxsize=16cm
$$
\epsfbox{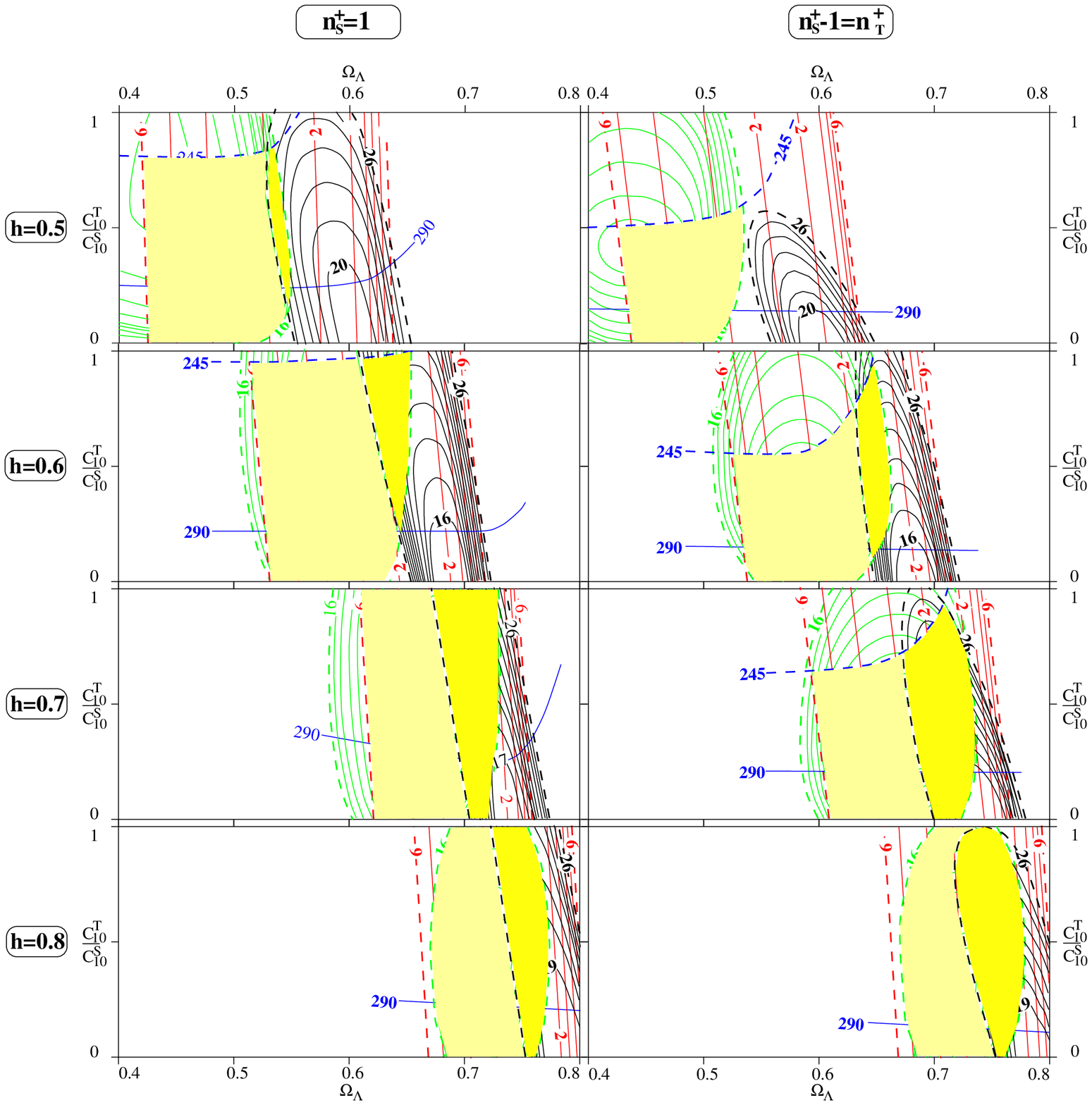}
$$
\end{figure*}

For all $h$ considered here, intersections with the previously preferred
regions are found, and enhanced in yellow (or gray) on Figure \ref{SIGW2}.
As expected from what was just said,
the major part of the preferred regions correspond to `inverted' steps
$0.45<p<1$. Those with $p=1$\ (Harrison-Zeldovich) and $1<p<1.1$\ (`usual' 
step) correspond to the lower left corners of the
preferred regions (not shown in Figure 2), with $C_{10}^T/C_{10}^S\leq 0.15$.

\subsection{Clusters distribution}

Let us now consider the rich Abell galaxy cluster distribution data
which constitutes the main motivations for the characteristic scale appearing
in our model, and which determines its location through the value of $k_0$.
The elliptic preferred regions (see Figure \ref{SIGW}) are narrow when
compared with STROMLO-APM regions, because data error bars are
smaller, and sensitive to larger scales (where $p$, and therefore
$C_{10}^T/C_{10}^S$\ values are crucial).
The best values are obtained in the absence
of gravitational waves, but values of $C_{10}^T / C_{10}^S \simeq 1$\ are still
acceptable (except for $n_S=1+n_T$, $h<0.6$).
There is no intersection with the CMB preferred region when
$h < 0.5$. The intersection of all preferred regions is coloured in bright
yellow (or dark gray) on Figure \ref{SIGW2}, showing that:
\begin{itemize}
\item
for $h\geq 0.7$, larger preferred region in the $C_{10}^T/C_{10}^S-
\Omega_{\Lambda}$ plane are obtained, with $0.68\leq\Omega_{\Lambda}
\leq 0.77$ and $0\leq C_{10}^T/C_{10}^S\leq 1$.
\item
for $h=0.6$, a rather narrow region survives, for which the GW
contribution is substantial, $\Omega_{\Lambda}\sim 0.63$ and
$0.2 \leq C_{10}^T / C_{10}^S \leq 1$.
\item
for $h=0.5$, a marginal intersection between the CMB and cluster favoured 
regions
is found for $\Omega_{\Lambda}\sim 0.53$.
This possibility is still acceptable, because we are rather severe in the
definition of the preferred windows.
\end{itemize}

In the previous discussion, we made almost no distinction between
$n_S=1$\ and $n_S=1+n_T$, since the results are quite similar in both case.
This shows that our predictions do not depend very much on the assumptions
on large scale tilts, provided that the consistency relation is satisfied.
In particular, the results would still hold for an arbitrary constant scalar
tilt in the range $1+n_T(k_0)<n_S<1$. The only difference is that the 
preferred parameter windows and the lowest $\chi^2$ regions
are systematically obtained for smaller $C_{10}^T/C_{10}^S$\ when
$n_S=1+n_T$, while the first case, with $n_S\approx 1$, 
admits slightly larger values of $C^T_{10}/C^S_{10}$.

\section{CONCLUSION}
The generation of a GW cosmological background is an essential prediction
of all inflationary models. Amplitude and statistics of this background can
be computed from first principles, as also their contribution to the CMB
temperature (and polarization) anisotropies.
However, only part of the inflationary models have a primordial
GW background sufficiently large in order to be actually measured, using 
either CMB measurements, or else gravitational-wave antennas for their direct 
detection.
An accurate measurement of the CMB angular temperature anisotropy and the
CMB polarization (especially on angular scales $\theta \ge 1^{\circ}$)
may result in the remarkable discovery of a primordial GW background
on cosmological scales which, of course, would be of great importance by 
itself, and would also provide a very strong argument for the inflationary 
scenario of the early Universe. 
The already measured height of the first
Doppler peak for multipoles $l=200 - 300$, in spite of all the indeterminacies,
is so high that it precludes a significant GW contribution to the multipole
dispersion $C_{10}$ 
(where $l=10$ is the characteristic multipole for the COBE data) 
if the initial power spectrum of adiabatic perturbations is scale-free. 
Shifting to a BSI initial spectrum helps to avoid this obstacle and to 
keep the possibility of finding the GW background.

In this paper, as in Paper I, we have considered the BSI CDM model with
a positive cosmological constant and we have confronted it with recent
observational data on CMB
temperature anisotropies, large-scale galaxy-galaxy correlations, peculiar
velocities of galaxies, and spatial correlations of rich Abell clusters.
Both new ingredients of this model, as compared to the standard CDM model
with a flat initial perturbation spectrum, namely the cosmological constant
and the particular form of the BSI initial spectrum, have been already 
introduced earlier to account for hitherto unexplained data, without 
reference to a primordial GW background. Now we have shown that this model,
with a slightly different choice of its parameters as compared to the one  
considered
in Paper I, admits a large GW background, too. A positive cosmological
constant is essential for this since the initial power spectrum admits
an {\it inverted} step, $p<1$ (i.e., more power on small vs. large scales), 
without
which the quantity $(C_T/C_S)_{10}$ characterizing the relative GW contribution
to $C_{10}$ would be small.

We have considered two subclasses of this model, differing by the behaviour
of the scalar (density) power spectrum in its approximately scale-free part far
below the breakpoint at $k=k_0$. In the first case, the spectral slope is
$n_S\approx 1$ in this region (while we may not assume neither $n_T\approx 0$,
nor neglect its scale dependence). In the second case, 
$n_T=n_S-1={\rm const}<0$.
For each class, we have investigated how large this GW background can be
while still leaving the model in good agreement with observations. 
The difference between the two cases is small, as expected.
We find that the best models
are located in essentially the same window of the cosmological parameters $h$
and $\Omega_{\Lambda}$: $h\simeq 0.7,~\Omega_{\Lambda}\simeq 0.7$ while the GW
contribution to the temperature fluctuations on large angular scales,
measured here by the parameter $\frac{C_{10}^T}{C_{10}^S}$, can be as large
as that of the scalar fluctuations. Note that this is still of course many
orders of magnitude below a value allowing their possible direct detection
by ground-based gravitational-wave antennas.

\section*{ACKNOWLEDGEMENTS}
A.S. acknowledges financial support by the Russian Foundation for Basic Research,
grant 96-02-17591, by the Russian research project ``Cosmomicrophysics''
and by the INTAS grant 93-493-ext.
J. L. thanks Joanne Baker for useful comments on CAT 2 results.
The authors also thank an anonymous referee for numerous comments on the paper.

\end{document}